# Prospects and merits of metal-clad semiconductor lasers from nearly UV to far IR

**Jacob B. Khurgin** [*]

*Johns Hopkins University, Baltimore MD 21218, USA*
[*]*jakek@jhu.edu*

**Abstract:** Using metal-clad (or plasmonic) waveguide structures in semiconductor lasers carries a promise of reduced size, threshold, and power consumption. This promise is put to a rigorous theoretical test, that takes into account increased waveguide loss, Auger recombination, and Purcell enhancement of spontaneous recombination. The conclusion is that purported benefits of metal waveguides are small to nonexistent for all the band-to-band and intersubband lasers operating from UV to Mid-IR range, with a prominent exception of far-IR and THz quantum cascade lasers. For these devices, however, metal waveguides already represent the state of the art, and the guiding mechanism in them has far more in common with a ubiquitous transmission line than with plasmonics.

Cascade 250.5403 Plasmonics

**References and links**

1. L. A. Coldren, S W Corzine, Diode Lasers and Photonic Integrated Circuits, Wiley, New York (1995)
2. S. Nakamura, M. Senoh, S. Nagahama, N. Iwasa, T Matsushita, T Mukai, "Blue InGaN-based laser diodes with an emission wavelength of 450 nm Appl. Phys. Lett., **76**(22), 22-24 (2000),
3. J. Faist, F Capasso, D. L. Sivco, C. Sirtori, A. L. Hutchinson, A. Y. Cho[1] "Quantum cascade laser", Science **264**, 553–556 (1994).
4. M. T. Hill and M. C. Gather, Advances in small lasers Nature Photonics, **8**, 808-816 (2014)
5. M. Stockman, "Nanoplasmonics: past, present, and glimpse into future," Opt. Express,19, 22029-22106 (2011).
6. R. F. Oulton, V. J. Sorger, T. Zentgraf, R.-M. Ma, G, Bartal, and X. Zhang, "Plasmon lasers at deep subwavelength scale," Nature, **461**, 629-632 (2009).
7. M. T. Hill, M. Marell, E. S. Leong, B. Smalbrugge, and C.-Z. Ning, "Lasing in metal-insulator-metal sub-wavelength plasmonic waveguides," Opt. Express, **17**, 11107-11112 (2009).
8. K. Ding, Z. C. Liu, L. J. Yin, M. T. Hill, M. J. H. Marell, P. J. van Veldhoven, R. Nöetzel, and C. Z. Ning, "Room-temperature continuous wave lasing in deep-subwavelength metallic cavities under electrical injection," Phys. Rev. B **85**, 041301(R) (2012).
9. S.-H. Kwon, J.-H. Kang, C. Seassal, S.-K. Kim, P. Regreny, Y.-H. Lee, C. M. Lieber, and H.-G. Park, "Subwavelength Plasmonic Lasing from a Semiconductor Nanodisk with Silver Nanopan Cavity," Nano Lett. **10**, 3679-3683 (2010).
10. A. Lakhani, M.-K. Kim, E. K. Lau, and M. C. Wu, "Plasmonic crystal defect nanolaser," Opt. Express **19**, 18237-18245 (2011).
11. J. H. Lee, M. Khajavikhan, A. Simic, and Q. Gu, "Electrically pumped sub-wavelength metallo-dielectric pedestal pillar lasers," Opt. Express **19**, 21524-21531 (2011).
12. M. P. Nezhad, A. Simic, O. Bondarenko, B. Slutsky, A. Mizrahi, L. Feng, V. Lomakin and Y. Fainman, "Room-temperature subwavelength metallo-dielectric lasers," Nat. Photon, **4**, 395-399 (2010).
13. M. Khajavikhan, A. Simic, M. Katz, J. H. Lee, B. Slutsky, A. Mizrahi, V. Lomakin and Y. Fainman, "Thresholdless nanoscale coaxial lasers," Nature **482**, 204-207 (2012).
14. R. F. Oulton, "Plasmonics: Loss and gain," Nat. Photon. **6**, 219–221 (2012).
15. R. F. Oulton, "Surface plasmon lasers: sources of nanoscopic light," Mater. Today **15**, 26–34 (2012).
16. R.-M. Ma, R.F. Oulton, V. J. Sorger, and X. Zhang, "Plasmon lasers: coherent light source at molecular scales," Laser Photon. Rev. 1-21 (2012)
17. D. J. Bergman and M. I. Stockman, "Surface Plasmon Amplification by Stimulated Emission of Radiation: Quantum Generation of Coherent Surface Plasmons in Nanosystems," Phys. Rev. Lett. **90**, 027402 (2003).
18. M. I. Stockman, "Spasers explained," Nat. Photon. **2,** 327-329 (2008).
19. M. I. Stockman, "The spaser as a nanoscale quantum generator and ultrafast amplifier," J. Opt. **12** 024004 (2010).

20. J. B. Khurgin, G. Sun, "Injection pumped single mode surface plasmon generators: Threshold, linewidth, and coherence" Opt. Express 20, 15309–15325 (2012)
21. J. B. Khurgin, G. Sun, "Comparative analysis of spasers, vertical-cavity surface-emitting lasers and surface-plasmon emitting Diodes" Nature Photonics 8, 468–473 (2014)
22. J. B. Khurgin and G. Sun: "How small can " Nano " be in a " Nanolaser "?", Nanophotonics, **1**, 3-8 (2012)
23. D. Wu, H. Wang, B. Wu, H. Ni, S. Huang, Y. Xiong, P. Wang, Q. Han, Z. Niu, I. Tangring and S.M. Wang, "Low threshold current density 1.3 mm metamorphic InGaAs/GaAs quantum well laser diodes", Electronics Letters, 44 (7). 474-475 (2008)
24. Y. Yao, A. J. Hoffman and C. F. Gmachl, " Mid-infrared quantum cascade lasers", Nature Photonics, 6, 432-429 (2012)
25. K. Ohtani, M. Beck, and J. Faist, "Double metal waveguide InGaAs/AlInAs quantum cascade lasers emitting at 24μm" Appl. Phys. Lett 105, 121115 (2014)
26. B. S. Williams, S. Kumar, H. Callebaut, ,Q. Hu, J. L. Reno "Terahertz quantum-cascade laser at 100 mm using metal waveguide for mode confinement"Appl. Phys. Lett 83, 2124-2126 (2003)
27. M. A. Belkin, J. A. Fan, S. Hormoz.F. Capasso, "Terahertz quantum cascade lasers with copper metal-metal waveguides operating up to 178 K", Opt. Express, 16, 3242-3248 (2008)
28. M. Martl, J. Darmo, C. Deutsch, M. Brandstetter, A. Andrews, P. Klang, G. Strasser, and K. Unterrainer, "Gain and losses in THz quantum cascade laser with metal-metal waveguide," Opt. Express 19, 733-738 (2011).
29. C. Walther, G. Scalari M. I. Amanti, M. Beck, J. Faist , Microcavity laser oscillating in a circuit-based resonator, Science, 327, 1495-7 (2010)

**1. Introduction**

In the last half century semiconductor lasers (SL) have come a long way from being laboratory curiosity to becoming indispensable in every walk of life [1]. The salient features of SL's – compact size, high efficiency and ability to be modulated at high speed assure that the range of SL applications expands in step with the expansion of the range of wavelengths in which SL's can operate. Last decade has seen expansion of SL's range to near UV (nitride lasers) [2] and mid and far-IR (quantum cascade lasers) [3] and has seen the birth of photonic integrated circuits (PIC's) in which SL's serve as key components. Photonic integration places even more stringent demands on size and power consumption of SL's and the field has made significant strides in reduction of these two characteristics, which includes VCSEL's, micro resonator lasers, and use of quantum dot active materials. Yet in the end, all of these techniques are bounded by the diffraction limit hence there has been an extensive effort to circumvent the diffraction limit by various techniques, typically involving use of metals [4]. These efforts coincided with rapid advancements in the field of plasmonics [5], where the field concentration on the sub-wavelength scale have been achieved and successfully used to enhance various linear and nonlinear optical processes. For these reasons, any laser structure incorporating metal had become known as a "plasmonic", or, better "nano-plasmonic laser" [6-16]. Even a new term "Spaser" [17-19] had been coined to describe the generator of coherent surface plasmon polaritons. The validity of using the term "plasmonic" versus less trendy "metal clad" will be discussed further on, but one can summarize the practical developments by saying that metal-clad lasers with sub-wavelength confinement in one or two dimensions had been successfully demonstrated in various spectral regions with varied results. The lasing thresholds ranged from comparatively low in the far-IR and THz regions to prohibitively high in near IR and visible so that in the visible range the lasing could only be achieved by strong optical pumping. As far as lasing in truly sub-wavelength in all three dimensions structures, there has been plenty of theoretical research with only a single report of optically pumped lasing. In our prior work [20-22] we have addressed the issues impeding development of sub-wavelength injection pumped lasers, namely high loss in the metal and rapid increase in spontaneous recombination rate caused by Purcell factor. Most of our work had been focused on the aforementioned sub−λ in all 3 dimension lasers (or spasers) which were shown to have prohibitively high threshold currents, broad linewidth, ,high noise and low efficiency. Less attention had been paid to far more practical metal-clad devices which are many wavelengths long along lateral dimension, and, while perhaps not as exhilarating as spasers, have been operated by a number of groups in various spectral ranges. In this work we shall examine the performance of metal-clad SL's over a wide range of wavelengths – from 0.5μm to 100 μm using essentially a single criterion of whether use of metal waveguide to confine the laser mode beyond diffraction limit can lead to the substantial reduction of lasing threshold. Our results will show that delicate interplay of many factors: loss in the metal, free carrier loss in semiconductor, Auger recombination, Purcell enhancement of radiative rate leads to dramatically different results in different spectral regions.

## 2. Surface plasmon polaritons and metal-clad waveguides

As mentioned in introduction, when metal, or other material with a negative dielectric constant $\varepsilon$ is present, diffraction limit can be circumvented. The photons in the dielectric with refractive index $n_d$ that propagate with high velocity $c/n_d$ (and therefore having long wavelength $\lambda/n_d$) get coupled with electrons that move with orders-of-magnitude lower Fermi velocity and have correspondingly much short wavelength. The net result is a surface-plasmon polariton (SPP), a quasi-particle that is partially a photon (i.e. electromagnetic field) and partially a plasmon (collective charge oscillation of free electrons) that effectively has a wavelength that is shorter than $\lambda/n$ and can be confined to the dimensions beyond the diffraction limit. SPP's exist in two varieties – propagating ones that are the subject of this work and the localized ones. Localized SPP's can also be used as cavities of nano-lasers, but, as shown in [22] the lasers based on such structures with only a few, or a single lasing mode (spasers) have exceptionally high thresholds and low coherence, and is not amendable to injection pumping.

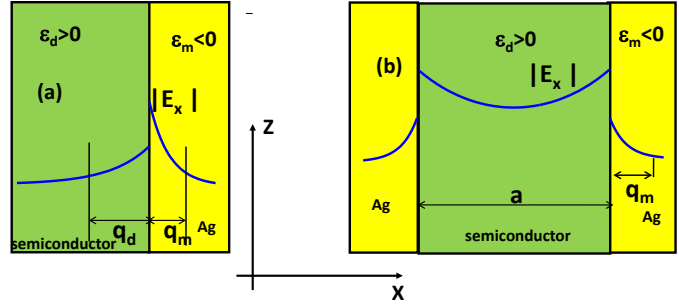

Fig.1 (a) Interface SPP and (b) gap SPP propagating in metal clad semiconductor waveguide.

The propagating SPP shown in Fig.1.a is essentially a transverse magnetic (TM) wave propagating on the interface between a dielectric with a positive dielectric constant $\varepsilon_d = n_d^2$ and a metal with a dielectric constant $\varepsilon_m = 1 - \omega_p^2/(\omega^2 + i\omega\gamma)$ whose real part is negative. The metal dielectric constant depends on two important factors – plasma frequency $\omega_p$ band momentum scattering rate $\gamma$. For most commonly used in plasmonics gold and silver plasma frequency is in deep UV ($\omega_p \sim 10^{16} s^{-1}$) while the scattering rate is on the scale of $\gamma > 0.5 \times 10^{14} s^{-1}$. The scattering rate is high because the density of states in the metal is extremely high in comparison to density of state of photons. The transverse electric field in the propagating SPP is

$$E_x = \begin{cases} \dfrac{\varepsilon_d}{\varepsilon_m} E_0 e^{q_m x} e^{j\beta z} & x > 0 \\ E_0 e^{-q_d x} e^{j\beta z} & x > 0 \end{cases} \quad (1)$$

If one introduces effective refractive index as the ratio of propagating constant $\beta$ to the wavevector in the dielectric, it can be found as

$$n_{eff} = \frac{\beta c}{\omega n_d} = \sqrt{\frac{\varepsilon_m}{\varepsilon_m + \varepsilon_d}} > 1 \;, \qquad (2)$$

while the decay constants in the metal and dielectric, also normalized to the wavevector in dielectric become $q_{m,eff} = \sqrt{n_{eff}^2 - \varepsilon_m/\varepsilon_d}$ and $q_{d,eff} = \sqrt{n_{eff}^2 - 1}$ respectively. The penetration length of SPP in the metal, is always very small, and the effective width of SPP can be defined as width that contains 95% of SPP energy, $W_{spp} \approx (3\lambda/4\pi n_d) q_{m,eff}^{-1}$. The minimum optical mode size achievable in all dielectric waveguides is about $\lambda/2n_d$, hence, if we define onset of the "deep sub-wavelength confinement" as $W_{spp} \leq \lambda/4n_d$, this confinement takes place when $n_{eff} \geq 1.4$, or $\omega \geq \omega_p/\sqrt{2\varepsilon_d + 1}$, which for silver and gold indicate $\lambda < 600nm$. Therefore using simple interface SPP can shrink the size of the propagating mode by a factor of two or narrower than the width of the mode in the dielectric waveguide only for the blue and green lasers. Gold becomes very lossy at these short wavelengths due to onset of interband absorption, which leaves one with silver that is not compatible with most SL fabrication processes. Therefore, one should look beyond a simple interface SPP for a scheme where deep-sub-wavelength confinement is attainable for any wavelength.

This bring us to the SPP's in metal slot waveguide, also referred to as the gap SPP shown in Fig. 1b, which of course has uncanny resemblance with the transverse electro-magnetic (TEM) wave in the microstrip ransmission line. The transverse electric field distribution in the waveguide with the slot width a is

$$E_x = \begin{cases} \frac{\varepsilon_d}{\varepsilon_m} E_0 e^{-q_m|x-a/2|} e^{j\beta z} & |x| > a/2 \\ E_0 \cosh(q_d x) & |x| < a/2 \end{cases} \qquad (3)$$

and obviously the effective width of the SPP mode is equal to the slot width no matter what is the frequency. The normalized decay constant can be found from the self-consistent dispersive relation

$$q_{d,eff} = -\frac{\sqrt{(1 - \varepsilon_m/\varepsilon_d) + q_{d,eff}^2}}{(\varepsilon_m/\varepsilon_d) \tanh(q_{d,eff} \pi n_d a/\lambda)} \qquad (4)$$

For the frequencies far from the surface plasmon resonance, $\omega \ll \omega_p/\sqrt{\varepsilon_d + 1}$, or practically speaking for $\lambda > 800nm$, $|\varepsilon_m| \gg \varepsilon_d$ which makes $q_{d,eff} \ll 1$, and for small gap size $a \ll \lambda$ one immediately obtains from (4)

$$q_{d,eff}^2 \approx -\frac{1}{(-\varepsilon_m/\varepsilon_d)^{1/2} (\pi n_d a/\lambda)} \approx \frac{\lambda_p}{a} \qquad (5)$$

and $n_{eff} \approx \sqrt{1 + \lambda_p^2/a^2}$, indicating that effective index depends only on the ratio of the gap width and plasma wavelength $\lambda_p \approx 140nm$. This is a rather interesting result that indicates that independent of the wavelength, the electromagnetic wave propagating in slot waveguide retains its strictly photon character for as long as the slot width is much larger than plasma wavelength and a very small fraction of energy penetrates the metal. Only for the slot sizes of less than a few hundred nanometers the energy is coupled into the oscillating motion of free electrons in the metal, and the mode can be called a "true" slot SPP, rather than a simple TEM wave.

As the field penetrates metal the loss of SPP mode increases and this loss can be estimated with a surprising ease using energy balance considerations. If one considers an electromagnetic mode in the dielectric waveguide, propagating with a wavevector $k_d = n_d \omega/c$ amplitudes of electric and magnetic fields at each point are related as $H = (\varepsilon/\mu)^{1/2} E$ and the densities of electric $u_E = \frac{1}{4}\varepsilon E^2$ and magnetic $u_H = \frac{1}{4}\mu H^2$ energies are equal to each other. Thus the energy oscillates back and forth between magnetic and electric forms, similar to the energy transfer between kinetic and potential energy in a mechanical oscillator. In the SPP with propagation constant $\beta = n_{eff} k_d$ the relation between the fields is $H = (\varepsilon/\mu)^{1/2} E/n_{eff}$ and the magnetic energy becomes less than electric energy, $u_H = u_E/n_{eff}^2$. The difference, $u_K = (1 - n_{eff}^{-2}) u_E$ is the kinetic energy of electrons in the metal. In the SPP the energy gets transferred back and forth between electric energy and combination of magnetic and kinetic energies, with the fraction of latter increasing with increase of $n_{eff}$. Since the electrons in the metal scatter with the rate $\gamma$ one can estimate the rate of energy loss of the SPP as $\gamma_{eff} \approx \gamma(1 - n_{eff}^{-2}) = \gamma q_{d,eff}^2/(1 + q_{d,eff}^2)$. For the wavelengths far from SP resonance one than can use (5) to obtain a result $\gamma_{eff} \approx \gamma \lambda_p/a$, from which the propagation length can be found as $L_{SPP} = \gamma_{eff}^{-1} v_g \approx \gamma^{-1} \lambda_p c/a n_d n_g$ where $v_g = \partial\omega/\partial\beta$ is a group velocity and $n_g = n_{eff} + \omega \partial n_{eff}/\partial\omega$ is effective group index. The result is most interesting as it shows that in this rough approximation SPP propagation length does not depend on the wavelength $\lambda$ This can be explained by the fact that imaginary part of the dielectric constant of the metal is roughly proportional to $\lambda$ while the propagation constant is proportional to $\lambda^{-1}$. It follows then, that for the long wavelengths the waveguide that is substantially sub-$\lambda$, still remains relatively large in comparison to the plasma wavelength and thus experiences lower loss. For example if we chose a quarter-wave wide gap $a = \lambda/4n_d$ we obtain $L_{SPP} \approx 4\gamma^{-1} c \lambda_p/\lambda n_g$. In addition to this, as the wavelength increases the inherent free carrier absorption $\alpha_{fc}$ in the doped regions of semiconductor, proportional to $\lambda^{-2}$ increases which makes metal loss relatively less important. All of this allows us to make a wide prediction that using metal clad waveguides in SL's is most advantageous at long, wavelengths, such as far IR and THz ranges.

### 3. Injection pumped metal clad SL's in the visible to near IR ranges

We now consider the possibility of achieving the lasing in the metal clad semiconductor waveguides using injection of carriers into –p-n diode. Rather than evaluating the lasing threshold which requires the gain to be sufficient to compensate the sum of waveguide loss and the mirror loss, which makes threshold dependent on particular laser design, we simply evaluate the transparency gain coefficient $g_{tr} = L_{SPP}^{-1} + \alpha_{fc}$, required to compensate the sum of losses in metal and doped semiconductor regions, transparency carrier density $N_{tr}$ at which this gain is achieved, and the transparency current density $J_{tr}$ required to maintain this current density. In well –designed laser operating with decent slope efficiency mirror (or output coupling) loss must exceed the waveguide loss, actual threshold values are typically a factor of a few higher than transparency values, which does not make a big difference for our order-of-magnitude comparison.

To find $N_{tr}$ one finds the frequency dependent gain at a given carrier density $N_{tr}$ as

$$g(\omega, N) = \frac{2\alpha_0 A}{3 n_d n_{eff}} \left( \frac{2\mu_r}{\hbar^2} (\hbar\omega - E_g) \right)^{1/2} [f_c(E_c, N) - f_v(E_v, N)] \qquad (6)$$

where $\alpha_0 = 1/137$ is a fine structure constant, $\mu_r$ is a joint density of state mass, $f_c(E_c, N)$ and $f_v(E_v, N)$ are the Fermi function of the conduction and valence band states involved in the optical transition with energy $\hbar\omega = E_c - E_v$, $E_g$ is the bandgap energy, and $A = 2\mu_r P^2 / m_0^2 E_g$ where $P$ is the transition matrix element and $m_0$ is the free electron mass. As a consequence of k.p theory of band structure near the zone center, for most of the semiconductors considered here $A$ is close to unity, ranging from 1.1 for GaAs to 1.3 for InAs.

Equating the maximum gain from (6) to the transparency gain $g_{tr}$ yields the value of transparency carrier concentration $N_{tr}$ and the value of transparency current density is found as the sum of radiative recombination and Auger recombination currents, i.e. $J_{tr} = ea\left[R_{rad}(N_{tr}) + CN_{tr}^3\right]$, where $C$ is Auger recombination coefficient, and the radiative recombination rate is

$$R_{rad} = \frac{8\alpha_0 A n_r}{3\hbar\lambda^2} \left(\frac{2\mu_r}{\hbar^2}\right)^{1/2} \int F_P(\omega) f_c (1-f_v) \left(\hbar\omega - E_g\right)^{1/2} d(\hbar\omega) \qquad (7)$$

where Purcell factor, caused by increased density of states is

$$F_P(\omega) = 1 + \frac{\lambda n_{eff} n_g}{8\pi a n_d} \qquad (8)$$

Let us now consider the results for the Ag waveguides with three different active media, In$_{0.4}$Ga$_{0.6}$N laser emitting in blue-green at $\lambda = 490 nm$, GaAs laser emitting in near IR $\lambda = 820 nm$, and In$_{0.53}$Ga$_{0.47}$As emitting at telecommunication wavelength $\lambda = 1550 nm$. In these calculations actual values of Ag dielectric constant are used, rather than Drude approximation.

In Fig.2a the dependence of the effective index $n_{eff}$ on gap width $a$ is shown – as explained in the previous section the dependence is generally flat for $a > \lambda_p \sim 140 nm$ and then rises rapidly. Similarly, transparency gain coefficient $g_{tr}$ shown in Fig.2b changes rather slowly for $a > \lambda_p$ and then experiences drastic increase as field penetrates inside the metal. Both effective index and transparency gain are higher for shorter wavelengths as dielectric constant of Ag deviates from the Drude expression and also increase in carrier-carrier scattering rate at higher photon energies. The effective propagation length of SPP $L_{SPP} \approx g_{tr}^{-1}$ ranges from tens of micrometers at $\lambda = 1550 nm$ to just a few micrometers at $\lambda = 490 nm$.

In Fig.2.c Purcell factor $F_P$ is shown which becomes important only at rather narrow gap width of 50nm. Transparency carrier density $N_{tr}$ (Fig.2d) ranges from few times $10^{18} cm^{-3}$ for wide In$_{0.53}$Ga$_{0.47}$As waveguides to as high as $10^{20} cm^{-3}$ in the narrow In$_{0.4}$Ga$_{0.6}$N structures. The ultimate result is the dependence of the transparence current $J_{tr}$ shown in Fig.2.e. As one can see $J_{tr}$ is high, yet reasonable 3-5 kA/cm$^2$ for In$_{0.53}$Ga$_{0.47}$As waveguides that are more than 200nm wide. But these are the transparency and not threshold currents which should be in the range 10kA/cm$^2$ which is more than an order of magnitude higher than what can be achieved in the conventional double heterostructure lasers with the comparable mode size [23]. As the gap size decreases the transparency gain increases and reaches 100kA/cm$^2$ at a~50nm when the effective index reaches the value of $n_{eff} \sim 1.4$, i..e when the wave propagating in the gap can be considered a "true" SPP, and the confinement is "deep sub-wavelength". Clearly, increase in the metal loss and onset of Auger recombination increase the threshold current density far more than it is reduced by the reduction of active volume. Situation is more dire for shorter wavelengths, where the $J_{tr}$ is always at least a few tens of kA/cm$^2$. Interestingly, at very small gap dimensions the transparency current in GaAs exceeds that in InGaN due to Auger recombination. Also notice that Purcell factor here brings no benefit whatsoever because it increases the density of all modes into which spontaneous decay occurs, while the stimulated emission takes place into just one propagating mode. From the point of reaching transparency, the only benefit of having a smaller gap is reduction of active volume size, but this benefit happens to be outweighed by the increase in loss as the field penetrates the metal. Furthermore, active volume in conventional SL's is also routinely decreased by using quantum well and quantum dot lasers with separate confinement and in this structures the transparency current density is indeed reduced dramatically. Note also that high pump densities, reaching 1 MW/cm$^2$ and more can be achieved by pulsed optical pumping [14-16], which is remarkable from the scientific point of view, but, practical applicability of these devices is debatable.

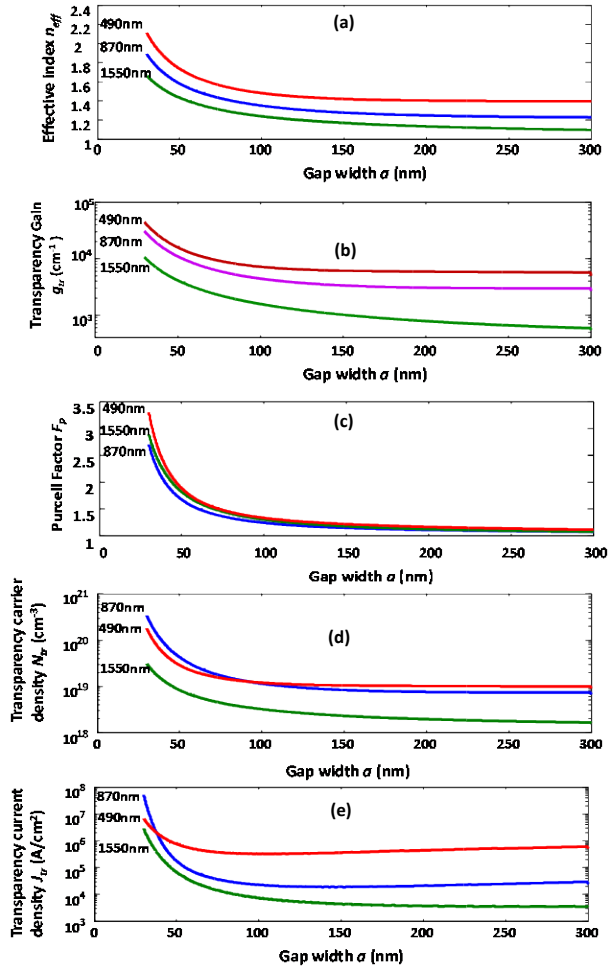

Fig.2 Dependences of pertinent characteristics on metal clad injection SL's operating at different wavelength in visible and near IR on the gap size $a$. (a) Effective index (b) Transparency gain coefficient (c) Purcell factor (d) Transparency carrier density and (e) Transparency current density.

## 4. Metal clad quantum cascade lasers

We now turn our attention to the long wavelength quantum cascade lasers (QCL's) [3] based on inter-subband transitions in semiconductor superlattices that can be designed to operate anywhere from mid-IR to THz range. Using metal clad waveguides in the IR and THz ranges may be advantageous for more than one reason. First of all, as operational wavelength $\lambda$ extends beyond 10 μm, the minimum thickness of the active layer $\lambda/2n_d$ becomes progressively thicker .Second, the index contrast between lattice matched III-V semiconductors decreases with the increase of wavelength, which adversely affects the confinement and further increases the required thickness of active region beyond $\lambda/2n_d$. Third, as mentioned above, the free carrier losses in the doped semiconductors increase as $\lambda^2$, while at which makes metal loss relatively less important. And finally, the metal loss itself decreases with wavelength as smaller fraction of the field is contained in the metal.

We have analyzed three different QCL's operating in the mid-IR at $\lambda = 4.5\mu m$ with parameters from [24], in far-IR $\lambda = 24\mu m$ (described in [25]) , and THz QCL at $\lambda = 100\mu m$ as described in [26]. Since the losses of Ag and Au at long wavelengths are roughly equal, we have used Au in our numerical calculations, results of which are shown in Fig.3

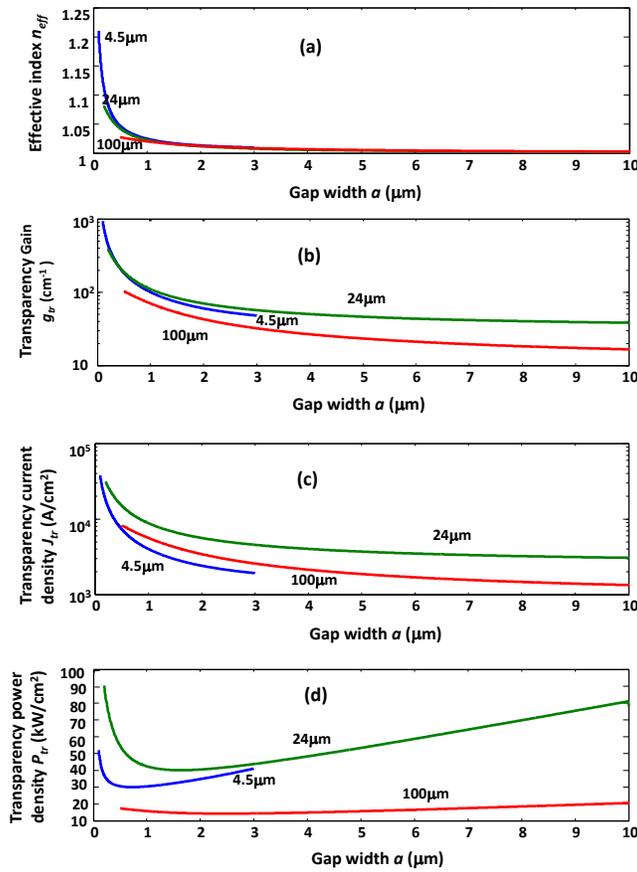

Fig.3 Dependences of pertinent characteristics on metal clad QCL's operating at different wavelengths from mid IR to THz on the gap size *a*. (a) Effective index (b) Transparency gain coefficient (c) Transparency current density (d) Transparency power density

In Fig3a one can see the change of effective index as a function of gap width. Since in mid and far IR dielectric properties of Au are well described by Drude formula, as expected from our analysis the three curves for three very different wavelengths are nearly identical. Furthermore, the field penetrates the metal and effective index exceeds unity only for sub-micron gap sizes, hence one can expect the metal losses to stay relatively low. As shown in Fig3b the transparency gain coefficients are indeed much lower than in near IR and visible, but quite different for three different wavelengths. This has to do with different free carrier absorption as well as with strong lattice absorption at $\lambda = 24 \mu m$ which is not far from the Restrahlen region.

Transparency current densities are shown in Fig3.c – they are lower than in near IR and visible, but for the mid IR wavelength $\lambda = 4.5 \mu m$ the current densities still exceed those reported with all dielectric waveguides in [24] which shows that in mid-IR metal clad waveguides are still inferior to the dielectric ones. For the longer far IR and THz wavelengths the metal clad waveguide becomes a necessity, and these waveguides were indeed successfully used in the works [25,26] as well as in many others [27,28] A strong argument for using metal clad waveguides can be seen in Fig.3d showing that as the active layer thickness is reduced, so is the voltage drop on QCL and the power density required for the transparency.

## 5. Conclusions

In this work we have analyzed the impact of using metal clad cavities to reduce size and threshold power density of electrically-pumped SL's operating in wide range of wavelengths – from blue-green to 100 μm. In order not to be tied to any particular laser design we have concentrated on evaluation of the transparency current and power densities required to compensate for all the intrinsic waveguide losses, dominated by the loss in the metal. Our conclusions can be summarized as follows. First of all, realistically, loss compensation can be achieved only in the waveguides that are wide enough to keep the effective wavelength not much shorter than wavelength in the dielectric (effective index only slightly higher than unity). Hence the actual laser will have to be at least half wavelength long in the longitudinal direction. The wave propagating in such structure retains its photon character with a very small contribution from the plasmons, and, in fact has much more in common with ubiquitous microwave transmission line than with SPP. If anything, characterizing such laser as a "spaser" would be preposterous. Second, on the practical level, up to and including mid-IR region, realistic metal clad lasers have substantially higher thresholds than all-dielectric counterparts, while actually not being much smaller than the latter. Only in the far-IR and THz regions metal clad lasers do hold significant advantage, the fact well known and universally used, but once again, in these regions, talking about plasmonics makes little sense since the frequencies become comparable or less than scattering rates in the metal, making the whole concept of plasmon questionable. Using analogy with electronic circuits, so elegantly done in [29] makes far more sense.

**Acknowledgement**

The author acknowledges steadfast backing by MIRTHE (NSF-ERC)
.